\documentclass{ws-p8-50x6-00}

\usepackage{url}
\newcommand{\ds}{{\sffamily DarkSUSY}}
\newcommand{\dsver}{3.14.01-beta}

\newcommand{\cpc}[3]{Comm.\ Phys.\ Comm.\ {\bf #1} (#2) #3}
\newcommand{\prl}[3]{Phys.\ Rev.\ Lett. {\bf #1} (#2) #3}

\newcommand{\prd}[3]{Phys.\ Rev.\ {\bf D#1} (#2) #3}

\newcommand{\href}[2]{#1}

\def\rn{\noindent\parshape 2 0truecm 8.5truecm 0.3truecm 8.2truecm}
\def\rn{}
\def\nn#1 #2{#2. #1}                            
\def\nnn#1 #2 #3{#2. #3. #1}                    
\def\nnnn#1 #2 #3 #4{#2. #3. #4. #1}            
\def\nnnnn#1 #2 #3 #4 #5{#2. #3. #4. #5. #1}    

\def\rf#1;#2;#3;#4;#5 {{\frenchspacing\par\rn#1, #3 {\bf #4}, #5 (#2). \par}}
\def\rfbook#1;#2;#3;#4;#5 {{\frenchspacing\par\rn#1, {\it #3} (#5, #4, #2).\par}}
\def\rfprep#1;#2;#3 {{\frenchspacing\rn#1, #3 (#2);\ }}
\def\rfprepend#1;#2;#3 {{\frenchspacing\rn#1, #3 (#2).}}

\begin{document}

\title{DarkSUSY -- A numerical package for dark matter
calculations in the MSSM}

\author{Paolo Gondolo\footnote{E-mail: pxg26@po.cwru.edu}}
\address{Department of Physics, Case Western Reserve University, 
10900 Euclid Ave., Cleveland OH 44106-7079, USA}

\author{\underline{Joakim Edsj\"o}\footnote{Rapporteur, E-mail: 
edsjo@physto.se} and Lars Bergstr\"om\footnote{E-mail: lbe@physto.se}}
\address{Department of Physics, Stockholm University, Box 6730, SE-113 
85 Stockholm, Sweden}

\author{Piero Ullio\footnote{E-mail: piero@tapir.caltech.edu}}
\address{Mail Code 130-33, California Institute of Technology,
Pasadena, CA 91125, USA}

\author{Edward A. Baltz\footnote{E-mail: eabaltz@physics.columbia.edu}}
\address{ISCAP
Columbia Astrophysics Laboratory, 550 W 120th St., Mail Code 5247
New York, NY 10027, USA}


\maketitle

\abstracts{The question of the nature of the dark matter in the
Universe remains one of the most outstanding unsolved problems in
basic science.  One of the best motivated particle physics candidates
is the lightest supersymmetric particle, assumed to be the lightest
neutralino.  We here describe \ds, an advanced numerical {\sc FORTRAN}
package for supersymmetric dark matter calculations which we release
for public use.  With the help of this package, the masses and
compositions of various supersymmetric particles can be computed, for
given input parameters of the minimal supersymmetric extension of the
Standard Model (MSSM).  For the lightest neutralino, the relic density
is computed, using accurate methods which include the effects of
resonances, pair production thresholds and coannihilations. 
Accelerator bounds are checked to identify viable dark matter
candidates.  Finally, detection rates are computed for a variety of
detection methods, such as direct detection and indirect detection
through antiprotons, gamma-rays and positrons from the Galactic halo
or neutrinos from the center of the Earth or the Sun.}

\section{Introduction}

One of the favourite candidates for the dark matter is a Weakly
Interacting Massive Particle, a WIMP. In supersymmetric extensions of
the standard model, the neutralino emerges as a natural WIMP
candidate for the dark matter of the universe.
There is, however, not a unique way of extending the standard model with
supersymmetry, but it is a general practice to use the simplest
possible model, the minimal supersymmetric enlargement of the standard
model (the MSSM), usually with some additional simplifying
assumptions.  

Over several years, we have developed analytical and numerical {\sc
Fortran} tools for dealing with the sometimes quite complex
calculations necessary to go from given input parameters in the MSSM
to actual quantitative predictions of relic density in the universe of
the neutralinos, and the direct and indirect detection rates.  The
program package, which we have named \ds, has now reached such a level
of sophistication and maturity that we find the time appropriate for
its public release.  In the following sections, we briefly describe
the different components of \ds, and refer the reader to our upcoming
paper \cite{ds} for more details.

For download of the latest version of \ds\, 
please visit the official \ds\ website, 
{\tt http://www.physto.se/\~{}edsjo/darksusy/}.
The version of \ds\ described here is \dsver.

\section{Definition of the Supersymmetric model}

We work in the framework of the minimal supersymmetric extension of
the standard model defined by, besides the particle content and gauge
couplings required by supersymmetry, the superpotential and the soft
supersymmetry-breaking potential.  See \cite{ds,bg} for details.

It is obvious that the parameter space of the most general MSSM is
huge, being specified by 124 a priori free parameters \cite{dimo}.  In
\ds\ we reduce this set by assuming that the off-diagonal elements of
the trilinear matrices {\bf A} and the scalar mass matrices {\bf M} 
are zero, and imposing CP conservation (except in the CKM matrix).

We then calculate all masses and most of the vertices entering 
the Feynman rules, which are all available to the user.
We include several options for the loop corrections to the 
Higgs masses.\cite{feynhiggs1,drees92,carena95,carena96} We also 
include loop corrections to the neutralino and chargino 
masses.\cite{NeuLoop1,NeuLoop2}

\section{Accelerator bounds}

Accelerator bounds can be checked by a call to a subroutine.  By
modifying an option, the user can impose bounds as of different
moments in time.  The default option in version \dsver\ adopts the
2000 limits by the Particle Data Group \cite{PDG} modified slightly
for the Higgs masses.  The user is also free to use his own routine to
check for experimental bounds, in which case he or she would only need
to provide an interface to \ds.

\section{Calculation of the relic density}

In \ds, we calculate the relic density by using the full cross 
section, including all resonances and thresholds and solve the 
Boltzmann equation numerically with the method given in 
\cite{GondoloGelmini,coann}.

When other supersymmetric particles are close in mass to the
lightest neutralino they will also be present at the time of 
freeze-out in the early Universe.  When this happens so called
coannihilations can take place between all these supersymmetric
particles.  We include the coannihilation
processes between all charginos and neutralinos lighter than
$f_{co}m_{\chi}$.  The mass fraction parameter $f_{co}$ is by default
set to 2.1 or 1.4 depending on how the relic density routines are
called (high accuracy or fast calculation), but can be set to anything
by the user.  We have not included coannihilations with squarks and
staus which occurs more accidentally than the in many cases
unavoidable mass degeneracy between the lightest neutralinos and the
lightest chargino.

We have included all two-body final states that occur at tree level, 
both for neutralino-neutralino, neutralino-chargino and 
chargino-chargino annihilations. Annihilation to $gg$, $\gamma 
\gamma$ and $Z\gamma$ that occur at the 1-loop level 
are also included.

\section{Detection rates}

The different detection rates for neutralino dark matter have been
calculated by many authors in the past.  We will here only give a
brief review about what is included in \ds, and which calculations
they are based on.  For a more extensive list of references, we refer
to \cite{ds}.

\subsection{Halo models}

Currently implemented in \ds\ is the spherical family of halo 
profiles 
  $\rho(r) \propto 1/\left((r/a)^{\gamma}\;[1+(r/a)^{\alpha}]^
  {(\beta-\gamma)/\alpha}\right)$
where e.g.\ the Navarro, Frenk and White profile \cite{navarro} is
given by $(\alpha,\beta,\gamma)=(1,3,1)$ and the isothermal sphere is
given by $(\alpha,\beta,\gamma)=(2,2,0)$.  The velocity distribution
is assumed to be a standard isotropic gaussian distribution.

\subsection{Direct detection}\label{subs:direct}

These routines calculate the spin-dependent and spin-independent 
scattering cross sections on protons and neutrons assuming the quark 
contributions to the nucleon spin from \cite{SMC}. The older set of 
data from \cite{jaffe} is also available as an option, but the user 
can set their own values if they wish.

\subsection{Monte Carlo simulations}

In several of the indirect detection processes below we need to 
evaluate the yield of different particles per neutralino annihilation.
The hadronization and/or decay of the annihilation products are 
simulated with {\sc Pythia} \cite{pythia} and the results are tabulated. 
These tables are then used by \ds.

\subsection{Neutrinos from the Sun and Earth}

Neutralinos can accumulate in the Earth and the Sun where they can 
annihilate pair-wise producing high energy muon neutrinos. The 
branching ratios for different annihilation channels are calculated 
and the {\sc Pythia} simulations are used to evaluate the yield of 
neutrinos. Neutrino interactions in the Sun as well as the charged 
current neutrino-nucleon interaction near the detector are also 
simulated with {\sc Pythia}.

There are routines to calculate a) the neutrino flux, b) the
neutrino-to-muon conversion rate and c) the neutrino-induced muon flux
either differential in energy and angle or integrated within an
angular cone and above a given threshold.  The new population of
neutralinos in the solar system (arising from neutralinos that have
scattered in the outskirts of the Sun) as described in
\cite{dk1,dk2,dkpop} can optionally be included as well.

\subsection{Antiprotons from halo annihilations}

Neutralinos can also annihilate in the Milky Way halo producing e.g.\
antiprotons.  These propagate in the galaxy before reaching us.  We
have implemented the propagation method described in \cite{pbar}. 
Optionally, the antiproton fluxes can also be solar modulated with the
spherically symmetric model of \cite{GleesonAxford}.  There are also
other propagation models\cite{pbarother} available as options.  The
antiproton fluxes are given differential in energy.

\subsection{Positrons from halo annihilations}

In neutralino annihilations in the halo we can also produce positrons. The 
flux of positrons is calculated with the propagation model in 
\cite{baltz} (with two choices of the energy dependence of the 
diffusion constant). The model in \cite{kamturner} can also be used as an 
option. The positron fluxes are given differential in energy.

\subsection{Gamma rays from halo annihilations}

We can also produce gamma rays from the halo annihilations.  These are
either monochromatic, produced from 1-loop annihilation \cite{lp} into
$\gamma\gamma$ and \cite{ub} $Z\gamma$, or with a continuous
energy spectrum, produced from $\pi^{0}$ decays in quark jets\cite{beu}.

The flux of gamma rays can be obtained in any given direction on the
sky for the user's choice of $(\alpha, \beta, \gamma)$ in the halo
profile.  There are also routines to average the flux over a chosen
angular resolution.  The continuous gamma rays use {\sc Pythia}
simulations to calculate the gamma ray flux (differential in energy,
or integrated above an energy threshold).

\subsection{Neutrinos from halo annihilations}

We can also produce neutrinos from neutralino annihilations in the 
halo. Although the fluxes are small, there are routines to calculate 
a) the neutrino flux, b) the neutrino-to-muon conversion rate and 
c) the neutrino-induced muon flux either differential in energy and 
angle or integrated within an angular cone and above a given energy 
threshold.

\section{Conclusions}

Over the years we have developed this numerical package, \ds, for 
neutralino dark matter calculations in the Minimal Supersymmetric 
Standard Model, MSSM. The package is now publically released and 
available for download from \url{http://www.physto.se/~edsjo/darksusy}. 
We have here presented an overview of what the program can do, and 
refer the reader to the upcoming paper \cite{ds}, where the details 
will be given. A test program, provided with the distribution, shows 
in more detail how \ds\ is used.

\section*{Acknowledgments}
L.B.\ and J.E.\ thank the Natural Sciences Research Council for their 
support. We also thank the Aspen Center for Physics, where parts of 
this release were prepared.

\end{document}